\documentstyle[12pt]{article}

\begin{document}
\title{\bf {Casimir Effect for  Moving Branes in Static dS$_{4+1}$ Bulk }}
\author{M.R. Setare  \footnote{E-mail: rezakord@ipm.ir}
  \\{Physics Dept. Inst. for Studies in Theo. Physics and
Mathematics(IPM)}\\
{P. O. Box 19395-5531, Tehran, IRAN }}

\date{\small{\today}}
 \maketitle
\begin{abstract}
In this paper we study the Casimir effect for conformally coupled
massless scalar fields on background of Static dS$_{4+1}$
spacetime. We will consider the general plane--symmetric solutions
of the gravitational field equations and boundary conditions of
the Dirichlet type on the branes. Then we calculate the vacuum
energy-momentum tensor in a configuration in which the boundary
branes are moving by uniform proper acceleration in static de
Sitter background. Static de Sitter space is conformally related
to the Rindler space, as a result we can obtain vacuum expectation
values of energy-momentum tensor for conformally invariant field
in static de Sitter space from the corresponding Rindler
counterpart by the conformal transformation.

 \end{abstract}
\section{Introduction}
The Casimir effect is regarded as one of the most striking
manifestation of vacuum fluctuations in quantum field theory. The
presence of reflecting boundaries alters the zero-point modes of a
quantized field, and results in the shifts in the vacuum
expectation values of quantities quadratic in the field, such as
the energy density and stresses. In particular, vacuum forces
arise acting on constraining boundaries. The particular features
of these forces depend on the nature of the quantum field, the
type of spacetime manifold and its dimensionality, the boundary
geometries and the specific boundary conditions imposed on the
field. Since the original work by Casimir in 1948 \cite{Casi48}
many theoretical and experimental works have been done on this
problem (see, e.g.,
\cite{Most97,Plun86,Lamo99,Bord99,Bord01,Kirs01,Bord02,Milt02} and
references therein). There are several
  methods to calculate Casimir energy. For instance,  we can mention mode summation,
  Green's function method \cite{Plun86}, heat kernel method \cite{Kirs01}along
   with appropriate
   regularization schemes such as point separation \cite{chr},\cite{adler}
  dimensional regularization \cite{deser}, zeta function regularization
  \cite{{haw},{blu},{Remeo},{Elizalde},{by}}. Recently a general new methods to compute renormalized
   one--loop quantum energies and energy densities are given in
   \cite{{gram1},{gram2}} (see also \cite{eliz2}).\\
 The Casimir effect
can be viewed as a polarization of vacuum by boundary conditions.
Another type of vacuum polarization arises in the case of an
external gravitational fields \cite{Birrell,Grib94}. It is well
known that the vacuum state for an uniformly accelerated
observer, the Fulling--Rindler vacuum
\cite{Full73,Full77,Unru76,Boul75, sahrin}, turns out to be
inequivalent to that for an inertial observer, the familiar
Minkowski vacuum. Quantum field theory in accelerated systems
contains many of special features produced by a gravitational
field avoiding some of the difficulties entailed by
renormalization in a curved spacetime. In particular, near the
canonical horizon in the gravitational field, a static spacetime
may be regarded as a Rindler--like spacetime. Rindler space is
conformally related to the static de Sitter space and to the
Robertson--Walker space with negative spatial curvature. As a
result the expectation values of the energy--momentum tensor for a
conformally invariant field and for corresponding conformally
transformed boundaries on the de Sitter and Robertson--Walker
backgrounds can be derived from the corresponding Rindler
counterpart by the standard transformation  \cite{Birrell}. The
authors in \cite{Birrell} have been shown that the Minkowski
vacuum contains a thermal spectrum of Rindler particles. One can
also demonstrate this by showing that the Green functions in
Minkowski vacuum are Rindler thermal Green functions. In a similar
way one can relate the vacua of static de Sitter space and de
Sitter space have the same curvature, but static de Sitter space
is a member of Rindler class, while de Sitter space is a member of
Minkowski
space.\\
The past few years witnessed a growing interest among particle
physicists and cosmologists toward models with extra space-like
dimensions. This interest was initiated by string theorists
\cite{Witt96}, who exploited a moderately large size of an
external 11th dimension in order to reconcile the Planck and
string/GUT scales. Taking this idea further, it was shown that
large extra dimensions allow for a reduction of the fundamental
higher-dimensional gravitational scale down to the TeV-scale
\cite{Arka98}. An essential ingredient of such a scenario is the
confinement of the standard model fields on field theoretical
defects, so that only gravity can access the large extra
dimensions. These models are argued to make contact with an
intricate phenomenology, with a variety of consequences for
collider searches, low-energy precision measurements, rare decays
and astroparticle physics and cosmology.  An alternative solution
to the hierarchy problem was proposed in Ref. \cite{Rand99}. This
higher dimensional scenario is based on a non-factorizable
geometry which accounts for the ratio between the Planck scale and
weak scales without the need to introduce a large hierarchy
between fundamental Planck scale and the
compactification scale. The model consists of a spacetime with a single $%
S^1/Z_2$ orbifold extra dimension. In this context, the Casimir
energy arising between the two static boundaries has been computed
in \cite{Horawa,Peskin}, in the first of these two works, the
backreaction on the geometry was taken into account. The same
problem has been considered in five-dimensional anti-deSitter
space in \cite{Odintsov}. Soon, the generalization of an AdS, flat
or dS brane in the AdS bulk \cite{kach}, and of a flat or dS brane
in dS bulk were studied carefully \cite{ito1}. The localization of
gravity in these models has also been discussed \cite{ito2}. The
bulk Casimir effect for a conformal or massive scalar when the
bulk represents five-dimensional AdS or dS space with two or one
four-dimensional dS brane, has been considered in \cite{eliz1}
(see also {\cite{eliz3}-\cite{eliz6}}). The recently proposed
cyclic model of the universe \cite{TS} is also based on this
framework in which the motion and collision of two such branes is
responsible for the Big-Bang of the standard cosmology. \\ Recent
astronomical observations of supernovae and cosmic microwave
background \cite{Ries98} indicate that the universe is
accelerating
 and can be well approximated by a world with a positive cosmological constant. If the
  universe would accelerate indefinitely, the standard cosmology leads to an asymptotic dS
   universe. De Sitter spacetime plays an important role in the inflationary scenario,
   where an exponentially expanding approximately dS spacetime is employed to solve
    a number of problems in standard cosmology.
In this paper we are interested in studying the possible effects
of the Casimir energy in an scenario like the one mentioned before
in which two branes are moving by uniform acceleration through the
static de Sitter vacuum. The complete analysis of the problem is
in general too involved to obtain explicit analytic results and,
for that reason, we will consider a simplified model in which the
two branes are perfectly flat, ignoring possible gravitational
effects. In any realistic model of a brane collision process it
will be necessary to consider the  acceleration and the brane
curvature \cite{Rasanen}. To see similar model in which the two
branes are moving with constant relative velocity refer to \cite
{maro}, as the author of this refrence have been mentioned  "the
present analysis would be the first (velocity-dependent)
correction to the flat static case" then may be could say that our model is second (accelerated
-dependent) correction to the static case.\\
 This problem for the conformally coupled Dirichlet
and Neumann massless scalar and electromagnetic fields in four
dimensional Rindler spacetime was considered by Candelas and
Deutsch \cite{Cand77}. Investigation of local physical
characteristics in the Casimir effect, such as expectation value
of the energy-momentum tensor, is of considerable interest. In
addition to describing physical structure of the quantum field at
a given point, the energy-momentum tensor acts as the source in
the Einstein equations and therefore plays an important role in
modeling a self-consistent dynamics involving the gravitational
field. Here we will investigate the vacuum expectation values of
the energy-momentum tensor for the massless scalar field with
conformal curvature coupling and satisfying Dirichlet boundary
condition on the infinite plane in five spacetime dimension. Here
we use the results of Ref. \cite{sahrin} to generate vacuum
energy--momentum tensor for the static de Sitter background which
is conformally related to the Rindler spacetime. Previously this
method has been used in \cite{set5} to drive the vacuum stress on
parallel plates for scalar field with Dirichlet boundary
condition in de Sitter space. Also this method has been used in
\cite{set6} to derive the vacuum characteristics of the Casimir
configuration on background of conformally flat brane-world
geometries for massless scalar field
with Robin boundary condition on plates.\\

\section{Vacuum expectation values for the energy-momentum tensor}
In this paper we will consider a conformally coupled massless scalar field $%
\varphi (x)$ satisfying the equation
\begin{equation}
\left( \nabla _{\mu }\nabla ^{\mu }+\frac{3}{16} R\right) \varphi
(x)=0,  \label{fieldeq}
\end{equation}
on background of a dS$_{4+1}$ spacetime. In Eq. (\ref{fieldeq})
$\nabla _{\mu }$ is the operator of the covariant derivative, and
$R$ is the Ricci scalar for the corresponding metric $g_{ik}$. In
static coordinates $x^i=(t,r,\theta ,\theta _2,\phi )$, dS metric
has the form
\begin{equation}
ds_{{\rm dS}}^{2}=g_{ik}dx^idx^k=\left( 1-\frac{r^{2}}{\alpha
^{2}}\right) dt^{2}-\frac{dr^{2}}{1-\frac{r^{2}}{\alpha
^{2}}}-r^{2}d\Omega ^{2}_{3} , \label{ds2dS}
\end{equation}
where $d\Omega ^{2}_{3}$ is the line element on the
$3$--dimensional unit sphere in the Euclidean space, and the
parameter $\alpha $ defines the dS curvature radius. Note that
$R=12/\alpha ^2$. Our main interest in the present paper is to
investigate the vacuum
expectation values (VEV's) of the energy--momentum tensor for the field $%
\varphi (x)$ in the background of the above de Sitter spacetime
induced by two parallel plates moving with uniform proper
acceleration. we will consider the case of a scalar field
satisfying Dirichlet boundary condition on the surface of the
plates:
\begin{equation}
\varphi \mid _{\xi =\xi _{1}}=\varphi \mid _{\xi =\xi _{2}}=0.
\label{Dboundcond}
\end{equation}
The presence of boundaries modifies the spectrum of the
zero--point fluctuations compared to the case without boundaries.
This results in the shift in the VEV's of the physical quantities,
such as vacuum energy density and stresses. This is the well known
Casimir effect.\\
First of all let us present the dS line element in the form
conformally related to the Rindler spacetime. With this aim we
make the coordinate transformation $x^i\to x'^{i}=(\tau ,\xi
,{\mathbf{x}}' )$, ${\mathbf{x}}'=(x'^{2},x'^{3} ,x'^{4})$ (see
Ref. \cite{Birrell} for the case $3+1$-dimensional case)
\begin{eqnarray} \label{coord}
&& \tau =\frac{t}{\alpha },\quad \xi =\frac{\sqrt{\alpha ^2-r^2}
}{\Omega },\quad x'^{2}=\frac{r}{\Omega }\sin \theta \cos \theta
_2 ,\nonumber \\
&&    x'^{3}=\frac{r}{\Omega }\sin \theta \sin \theta _2\cos \phi,
\quad x'^{4}=\frac{r}{\Omega }\sin \theta \sin \theta _2 \sin
\theta _{2}\sin \phi,
\end{eqnarray}
with the notation
\begin{equation}\label{Omega}
  \Omega =1-\frac{r}{\alpha }\cos \theta .
\end{equation}
Under this coordinate transformation the dS line element takes the
form
\begin{equation}
ds_{{\rm dS}}^{2}=g'_{ik}dx'^idx'^k=\Omega ^2\left( \xi ^{2}d\tau
^{2}-d\xi ^{2}-d{\mathbf{x}}'^{2}\right) .  \label{ds2dS1}
\end{equation}
In this form the dS metric is manifestly conformally related to
the Rindler spacetime with the line element $ds_{{\rm R}}^{2}$:
\begin{equation}
ds_{{\rm dS}}^{2}=\Omega ^{2}ds_{{\rm R}}^{2},\quad ds_{{\rm
R}}^{2}=g_{{\mathrm{R}}ik}dx'^idx'^k=\xi ^{2}d\tau ^{2}-d\xi
^{2}-d{\mathbf{x}}'^{2},\quad g'_{ik}=\Omega ^2
g_{{\mathrm{R}}ik}. \label{confrel}
\end{equation}
 The Casimir effect with
boundary conditions (\ref{Dboundcond}) on two parallel plates
moving with uniform proper acceleration on background of the
Rindler spacetime is investigated in Ref. \cite{sahrin} for a
scalar field with a Dirichlet and Neumann boundary condition. The
expectation values of the energy-momentum tensor for a scalar
field $\varphi _{{\mathrm{R}}}(x')$ in the Fulling-Rindler vacuum
can be presented in the form of the sum
\begin{equation}
\left\langle 0_{R}|T_{i}^{k}[g_{{\mathrm{R}}lm},\varphi
_{{\mathrm{R}}}]|0_{R}\right\rangle =\left\langle
\tilde{0}_{R}|T_{i}^{k}[g_{{\mathrm{R}}lm},\varphi
_{{\mathrm{R}}}]|\tilde{0}_{R}\right\rangle +\left\langle
T_{i}^{k}[g_{{\mathrm{R}}lm},\varphi _{{\mathrm{R}}}]\right\rangle
^{(b)}, \label{TikR}
\end{equation}
where $|0_{R}\rangle $ are $|\tilde 0_{R}\rangle $ are the
amplitudes for the vacuum in the Rindler space in presence and
absence of the branes respectively, $\left\langle
T_{i}^{k}[g_{{\mathrm{R}}lm},\varphi _{{\mathrm{R}}}]\right\rangle
^{(b)}$ is the part of the vacuum energy-momentum tensor induced
by the branes. In the case of a conformally coupled massless
scalar field for the part without boundaries one has
\begin{equation}
\left\langle \tilde{0}_{R}|T_{i}^{k}[g_{{\mathrm{R}}lm},\varphi
_{{\mathrm{R}}}]|\tilde{0}_{R}\right\rangle =
\frac{\delta_{i}^{k}}{32\pi^{2}\xi^{5}}\int_{0}^{\infty}
\frac{\omega^{4}d\omega}{e^{2\pi\omega}+1}(\frac{1}{4\omega^{2}}+1).
\label{TikR0}
\end{equation}
For a scalar field $\varphi _{{\mathrm{R}}}(x')$, satisfying the
Dirichlet boundary condition, the boundary induced part in the
region between hypersurface have the form \cite{sahrin}
\begin{equation}
\left\langle T_{i}^{k}[g_{{\mathrm{R}}lm},\varphi
_{{\mathrm{R}}}]\right\rangle ^{(b)}= A_4 \delta
_{i}^{k}\int_{0}^{\infty }dkk^{4}\int_{0}^{\infty }d\omega
\,\left\{ \frac{\sinh \pi \omega }{\pi }\,f^{(i)}[\tilde{D}
_{i\omega }(k\xi ,k\xi _{2})] - \frac{I_{\omega }(k\xi
_{1})}{I_{\omega }(k\xi _{2})}\frac{F^{(i)}[D_{\omega }(k\xi ,k\xi _{2})]}{%
D_{\omega }(k\xi _{1},k\xi _{2})}\right\},  \label{TikRb}
\end{equation}
where
\begin{equation}\label{Adnot}
  A_4=\frac{1}{4\pi ^{5/2}\Gamma( 3/2)}.
\end{equation}
Also we have introduced the notation
\begin{equation}
\tilde{D}_{i\omega }(k\xi ,k\xi _{2})=K_{i\omega }(k\xi )-\frac{%
K_{i\omega }(k\xi _{2})}{I_{i\omega }(k\xi _{2})}I_{i\omega }(k\xi
), \label{ztilda}
\end{equation}
and the functions $F^{(i)}[G(z)]$, $i=0,...,4$ are as following
\begin{equation}
F^{(i)}[G(z)]=f^{(i)}[G(z),\omega \rightarrow i\omega ].
\label{Ffunc}
\end{equation}
Here for a given function $G(z)$ we use the notations
\begin{eqnarray}
f^{(0)}[G(z)] &=& \frac{1}{8} \left| \frac{dG(z)}{dz}%
\right| ^{2}+\frac{3
}{16z}\frac{d}{dz}|G(z)|^{2}+\frac{1}{8}\left[ 1 +7\frac{\omega
^{2}}{z^{2}} \right]
|G(z)|^{2},  \label{f0} \\
f^{(1)}[G(z)] &=&-\frac{1}{2}\left| \frac{dG(z)}{dz}\right|
^{2}-\frac{3
}{16z}\frac{d}{dz}|G(z)|^{2}+\frac{1}{2}\left( 1-\frac{\omega ^{2}}{z^{2}}%
\right) |G(z)|^{2},  \label{f1} \\
f^{(i)}[G(z)] &=&-\frac{|G(z)|^{2}}{3}+ \frac{1}{8}
\left[ \left| \frac{dG(z)}{dz}\right| ^{2}+\left( 1-\frac{\omega ^{2}}{z^{2}}%
\right) |G(z)|^{2}\right] ;\quad i=2,3,4  \label{f23}
\end{eqnarray}
where $G(z)=D_{i\omega }(z,k\xi _{2})$, which given by  following
expression,  and the indices 0,1 correspond to the coordinates
$\tau $, $\xi $ respectively,
\begin{equation}
D_{i\omega }(k\xi ,k\xi _{2})=I_{i\omega }(k\xi _{2})K_{i\omega
}(k\xi )-K_{i\omega }(k\xi _{2})I_{i\omega }(k\xi ).
\label{Deigfunc}
\end{equation}
To find the vacuum expectation values generated by the branes in
the dS$_{4+1}$ space, first we will consider the corresponding
quantities in the coordinates $(\tau ,\xi ,{\mathbf{x}}')$ with
the metric (\ref{ds2dS1}). The latters are found by using the
standard transformation formula for the conformally related
problems:
\begin{equation}
\left\langle 0_{{\mathrm{dS}}}|T_{i}^{k}\left[ g'_{lm},\varphi
\right] |0_{{\mathrm{dS}}}\right\rangle =\Omega ^{-5}\left\langle
0_{{\mathrm{R}}}|T_{i}^{k}\left[ g_{{\mathrm{R}}lm},\varphi
_{{\mathrm{R}}}\right] |0_{{\mathrm{R}}}\right\rangle
+\left\langle T_{i}^{k}\left[ g'_{lm},\varphi \right]
\right\rangle ^{(an)}, \label{conftransemt}
\end{equation}
where the second summand on the right is determined by the trace
anomaly and is related to the divergent part of the corresponding
effective action by the relation \cite{Birrell}
\begin{equation}
\left\langle T_{i}^{k}\left[ g'_{lm},\varphi \right]
\right\rangle ^{(an)}=2g'^{kl}\frac{\delta }{\delta g'^{il}(x)}W_{{\rm %
div}}[g'_{mn},\varphi ].  \label{gravemt}
\end{equation}
Note that in odd spacetime dimensions the conformal anomaly is
absent and the corresponding anomaly part vanishes:
\begin{equation}
\left\langle T_{i}^{k}\left[ g'_{lm},\varphi \right] \right\rangle
^{(an)}=0.  \label{gravemteven}
\end{equation}
The formulae given above allow us to present the dS vacuum
expectation values in the form similar to (\ref{TikR}):
\begin{equation}
\left\langle 0_{{\rm dS}}|T_{i}^{k}\left[ g'_{lm},\varphi
\right]|0_{{\rm dS}}\right\rangle =\left\langle \tilde{0}_{{\rm
dS}}|T_{i}^{k}\left[ g'_{lm},\varphi \right]|\tilde{0}_{{\rm
dS}}\right\rangle +\left\langle T_{i}^{k}\left[ g'_{lm},\varphi
\right]\right\rangle ^{(b)} , \label{TikdS}
\end{equation}
where $\left\langle \tilde{0}_{{\rm dS}}|T_{i}^{k}\left[
g'_{lm},\varphi \right]|\tilde{0}_{{\rm dS}}\right\rangle $ are
the vacuum expectation values in the dS space without boundaries
and the part $\left\langle T_{i}^{k}\left[ g'_{lm},\varphi
\right]\right\rangle ^{(b)}$ is induced by the branes. Conformally
transforming the Rindler results one finds
\begin{eqnarray}
\left\langle \tilde{0}_{{\rm dS}}|T_{i}^{k}\left[ g'_{lm},\varphi
\right]|\tilde{0}_{{\rm dS} }\right\rangle  &=&\Omega ^{-
5}\left\langle \tilde{0}_{{\mathrm{R}}}|T_{i}^{k}|\tilde{0}
_{{\mathrm{R}}}\right\rangle +\left\langle T_{i}^{k}\left[
g'_{lm},\varphi \right] \right\rangle ^{(an)},  \label{TikdS0} \\
\left\langle T_{i}^{k}\left[ g'_{lm},\varphi \right]\right\rangle
^{(b)} &=&\Omega ^{-5}\left\langle
T_{i}^{k}[g_{{\mathrm{R}}lm},\varphi _{{\mathrm{R}}}]\right\rangle
^{(b)}. \label{TikdSb}
\end{eqnarray}
Under the conformal transformation $g'_{ik}=\Omega ^{2}g
_{{\mathrm{R}}ik}$, the field $\varphi _{{\mathrm{R}}}$  will
change by the rule
\begin{equation}
\varphi (x')=\Omega ^{-3/2}\varphi _{{\mathrm{R}}}(x'),
\label{phicontr}
\end{equation}
where the conformal factor is given by expression (\ref{Omega}).
The vacuum expectation values of the energy-momentum tensor in
coordinates are obtained from expressions (\ref{TikdS0}) and
(\ref{TikdSb}) by the standard coordinate transformation formulae.
As before, we will present the corresponding components in the
form of the sum of purely dS and boundary parts:
\begin{equation}
\left\langle 0_{{\rm dS}}|T_{i}^{k}\left[ g_{lm},\varphi
\right]|0_{{\rm dS}}\right\rangle =\left\langle \tilde{0}_{{\rm
dS}}|T_{i}^{k}\left[ g_{lm},\varphi \right]|\tilde{0}_{{\rm
dS}}\right\rangle +\left\langle T_{i}^{k}\left[ g_{lm},\varphi
\right]\right\rangle ^{(b)} . \label{TikdS1}
\end{equation}
By using the relations (\ref{coord}) between the coordinates for
the purely dS part one finds
\begin{equation}\label{Tik0dSst}
\left\langle \tilde{0}_{{\rm dS}}|T_{i}^{k}\left[ g_{lm},\varphi
\right]|\tilde{0}_{{\rm dS}}\right\rangle = \frac{(\alpha
^2-r^2)^{-\frac{5}{2}}}{32\pi ^{2}\Gamma
(2)\xi^{5}}\int_{0}^{\infty}
\frac{\omega^{4}d\omega}{e^{2\pi\omega}+1}(\frac{1}{4\omega^{2}}+1){\mathrm{diag}}\left(
-1,\frac{1}{4},\frac{1}{4},\frac{1}{4} ,\frac{1}{4} \right) .
\end{equation}
This formula generalizes the result for $3+1$-dimension given, for
instance, in Ref. \cite{Birrell}. As the for boundary induced
energy-momentum tensor the spatial part is not isotropic, the
corresponding part in the coordinates $x^i$ is more complicated:
\begin{eqnarray}\label{Tik0dSstb}
\left\langle T_{i}^{k}\left[ g_{lm},\varphi \right]\right\rangle
^{(b)} &=& \Omega ^{-5}\left\langle T_{i}^{k}\left[
g_{{\mathrm{R}}lm},\varphi _{{\mathrm{R}}}\right]\right \rangle
^{(b)}, \quad i,k=0,3,4, \\
\left\langle T_{1}^{1}\left[ g_{lm},\varphi \right]\right\rangle
^{(b)} &=&\frac{(\cos \theta -r/\alpha )^2}{\Omega
^{7}}\left\langle T_{1}^{1}\left[ g_{{\mathrm{R}}lm},\varphi
_{{\mathrm{R}}}\right]\right\rangle ^{(b)} \nonumber \\
&& +\frac{1-r^2/\alpha ^2}{\Omega ^{5}} \sin ^2\theta \left\langle
T_{2}^{2}\left[ g_{{\mathrm{R}}lm},\varphi
_{{\mathrm{R}}}\right]\right\rangle ^{(b)} , \\
\left\langle T_{1}^{2}\left[ g_{lm},\varphi \right]\right\rangle
^{(b)} &=&\frac{(r/\alpha -\cos \theta )\sin \theta }{r\Omega
^{7}}\left\{ \left\langle T_{1}^{1}\left[
g_{{\mathrm{R}}lm},\varphi _{{\mathrm{R}}}\right]\right\rangle
^{(b)}-\left\langle T_{2}^{2}\left[ g_{{\mathrm{R}}lm},\varphi
_{{\mathrm{R}}}\right]\right\rangle ^{(b)}\right\} ,\\
\left\langle T_{2}^{2}\left[ g_{lm},\varphi \right]\right\rangle
^{(b)} &=&  \frac{ 1-r^2/\alpha ^2}{\Omega ^{7}} \sin ^2\theta
\left\langle T_{1}^{1}\left[ g_{{\mathrm{R}}lm},\varphi
_{{\mathrm{R}}}\right]\right\rangle ^{(b)}\nonumber \\
&& + \frac{(r/\alpha -\cos \theta  ^2)}{\Omega ^{7}} \left\langle
T_{2}^{2}\left[ g_{{\mathrm{R}}lm},\varphi
_{{\mathrm{R}}}\right]\right\rangle ^{(b)} ,
\end{eqnarray}
where the expressions for the components of the boundary induced
energy-momentum tensor in the Rindler spacetime are given by
formula (10)-(16). As we see the resulting energy-momentum tensor
is non-diagonal.\\
In the discussion above we have considered the vacuum
energy-momentum tensor of the bulk. For a scalar field on
manifolds with boundaries in addition to the bulk part the
  energy-momentum tensor contains a contribution located on the boundary.
For arbitrary bulk and boundary geometries the expression of the
surface energy-momentum tensor is given in Ref. \cite{Saha03emt}.
In the case of a conformally coupled scalar field the
transformation formula forthe surface energy-momentum tensor
under the conformal rescaling of the metric is the same as that
for the volume part. For our problem in this paper, the surface
       energy-momentum tensor is obtained from the corresponding Rindler counterpart by a
       way similar to that described above. The expression for the latter is given
       in Ref. \cite{Saha03emt}.

\section{Conclusion}
Over the last few years a lot of interest has been raised on the
possibility that our universe is a $3-$brane embedded in a higher
dimensional spacetime. Ordinary matter fields are assumed to live
on the brane while gravity propagates in the whole spacetime. The
main part of the work done in this direction refers to the branes
sitting at a prescribed point of an extra dimension. However, it
is tempting, even inspired by $D-p$-brane models, to consider that
the three-brane is somehow let to move in the bulk.\\
In the present paper we have investigated the Casimir effect for a
conformally coupled massless scalar field between two boundary
branes moving by uniform acceleration, on background of the
five-dimensional static de Sitter spacetime. We have assumed that
the scalar field satisfies Dirichlet boundary condition on the
branes. The static de Sitter spacetime is conformally related to
the Rindler spacetime, then the vacuum expectation values of the
energy-momentum tensor are derived from the corresponding Rindler
spacetime results by using the conformal properties of the
problem.  The vacuum expectation value of the energy-momentum
tensor
    for a brane in dS spacetime consists of two parts given in Eq.(21).
    The first one corresponds to the purely dS contribution when the boundary
    is absent. It is determined by formula (22), where the second
    term on the right is due to the trace anomaly and is zero for odd spacetime
    dimensions. The second part in the vacuum energy-momentum tensor is due to
    the imposition of boundary conditions on the fluctuating quantum field.
     The corresponding components are related to the vacuum energy-momentum tensor
     in the Rindler spacetime by Eqs. (27)--(30) and the Rindler
      tensor is given by formulae (10)--(13). Unlike to the purely dS part, the boundary induced part of the
     energy-momentum tensor is non-diagonal and depends on both dS static
    coordinates $r$ and $\theta $.


\begin{thebibliography}{99}
\bibitem{Casi48} H. B. G. Casimir, Proc. K. Ned. Akad. Wet. {\bf
51}, 793 (1948).

\bibitem{Most97}  V. M. Mostepanenko and N. N. Trunov, {\it The
Casimir Effect and Its Applications} (Clarendon, Oxford, 1997).

\bibitem{Plun86}  G. Plunien, B. Muller, and W. Greiner, Phys. Rep.
{\bf 134}, 87 (1986).

\bibitem{Lamo99} S. K. Lamoreaux, Am. J. Phys. {\bf 67}, 850 (1999).

\bibitem{Bord99} {\it The Casimir Effect. 50 Years
Later} edited by M. Bordag (World Scientific, Singapore, 1999).

\bibitem{Bord01} M. Bordag, U. Mohidden, and V. M. Mostepanenko,
Phys. Rep. {\bf 353}, 1 (2001).

\bibitem{Kirs01} K. Kirsten, {\it Spectral functions in Mathematics
and Physics}. CRC Press, Boca Raton, 2001.

\bibitem{Bord02} M. Bordag, ed., Proceedings of the Fifth Workshop
on Quantum Field Theory under the Influence of External
Conditions, Int. J. Mod. Phys. {\bf A17} (2002), No. 6\&7.

\bibitem{Milt02} K. A. Milton, {\it The Casimir Effect: Physical Manifestation
of Zero--Point Energy} (World Scientific, Singapore, 2002).
 \bibitem {chr}  S. M. Christensen, Phys. Rev. {\bf D14}, 2490(1976); 17,
 946,(1978).
 \bibitem {adler}S. L. Adler, J. Lieberman and Y. J. Ng,  Ann. Phy. (N.Y) 106,
 279,(1977).
 \bibitem {deser}S. Deser, M. J. Duff and C. J. Isham, Nucl. Phys {\bf B11}, 45
 (1976),see also D. M. capper and M. J. Duff, Nuovo Cimento
 {\bf 23A}, 173, (1974); Phys. Lett. {\bf 53A}, 361, (1975).
  \bibitem {haw} S. W. Hawking, Commun. Math. Phys. 55, 133(1977).
  \bibitem{blu}S. Blau, M. Visser and A. Wipf, Nucl. Phys. {\bf
  B310}, 163, (1988).
  \bibitem{Remeo}E. Elizalde, S. D. Odintsov, A. Romeo, A. A.
Bytsenko and S. Zerbini, Zeta Regularization Techniques with
Applications(World Scientific, Singapore, 1994).

\bibitem{Elizalde}E. Elizalde, Ten Physical Applications of
Spectral Zeta Functions, Lecture Notes in Physics (Springer
Verlag, Berlin, 1995).
\bibitem{by}A. A. Bytsenko, G. Cognola, E. Elizalde, V. Moretti
and S. Zerbini,  Analytic aspect of quantum fields (World
Scientific, Singapore, 2003).
\bibitem{gram1} N. Graham, R. L. Jaffe, V. Khemani, M. Quandt, M. Scandurra, H.
 Weigel, Nucl. Phys.{\bf B645}, 49, (2002).
 \bibitem{gram2}N. Graham, R. L. Jaffe, V. Khemani, M. Quandt, M. Scandurra, H.
 Weigel, hep-th/0207205.
 \bibitem{eliz2}E. Elizalde, J. Phys. {\bf A36}, L567, (2003).
 \bibitem{Birrell} N. D. Birrel and P. C. W. Davies, {\it Quantum Fields in
Curved Space} (Cambridge: Cambridge University Press, 1982).
\bibitem{Grib94} A. A. Grib, S. G. Mamayev, and V. M.
Mostepanenko, {\it Vacuum Quantum Effects in Strong Fields} (St.
Petersburg, 1994).

\bibitem{Full73}S. A. Fulling, Phys. Rev {\bf D7}, 2850, (1973).

\bibitem{Full77}S. A. Fulling, J. Phys. A: Math. Gen. {\bf 10},
917, (1977).

\bibitem{Unru76}W. G. Unruh, Phys. Rev. {\bf D14}, 870, (1976).

\bibitem{Boul75}D. G. Boulware, Phys. Rev. {\bf D11}, 1404,
(1975).
\bibitem{Witt96} E. Witten, Nucl. Phys. {\bf B471}, 135, (1996); P.
Horava and E. Witten, Nucl. Phys. {\bf B460}, 506, (1996); T.
Banks and M. Dine, Nucl. Phys. {\bf B479}, 173, (1996).
\bibitem{Arka98} N. Arkani-Hamed, S. Dimopoulos, and G. Dvali, Phys. Lett.
{\bf B429}, 263, (1998); Phys. Rev. {\bf D59}, 086004, (1999); I.
Antoniadis, N. Arkani-Hamed, S. Dimopoulos, and G. Dvali, Phys.
Lett. {\bf B436}, 257, (1998).
\bibitem{Rand99}L. Randall and R. Sundrum, Phys. Rev. Lett. {\bf 83},
3370, (1999).
\bibitem{Horawa}M. Fabinger and P. Ho\v{r}ava,  Nucl. Phys. {\bf B580},
243, (2000).
\bibitem{Peskin}E. A. Mirabelli and M. E. Peskin,
 Phys. Rev. {\bf D58}, 065002, (1998).
\bibitem{Odintsov}S. Nojiri, S. D. Odintsov and S. Zerbini,
{\it Class. Quant. Grav.} {\bf 17}, 4855, (2000); I. Brevik, K.
Milton, S. Nojiri and S. Odintsov,  Nucl. Phys. {\bf B599}, 305,
(2001).
\bibitem{kach}S. Kachru, M. Schulz and E. Silverstein, Phys. Rev. {\bf D62},
045021, (2000).
\bibitem{ito1}M. Ito, hep-th/0206153.
\bibitem{ito2}M. Ito, hep-th/0204113.
\bibitem{eliz1} E. Elizalde, S. Nojiri, S. D. Odintsov, S. Ogushi,
Phys. Rev. {\bf D67}, 063515, (2003).
\bibitem{eliz3}E. Elizalde, J. Quiroga Hurtado,  Mod. Phys. Lett. {\bf A19}, 29, (2004).
\bibitem{eliz4}E. Elizalde, J. E. Lidsey, S. Nojiri, S. D.
Odintsov, Phys. Lett. {\bf B574}, 1, (2003).
\bibitem{eliz5} G. Cognola, E. Elizalde, S. Zerbini,
hep-th/0312011.
\bibitem{eliz6}G. Cognola, E. Elizalde, S. Nojiri, S. D. Odintsov,
 S. Zerbini, hep-th/0312269.
\bibitem{TS} J. Khoury, B.A. Ovrut, P. J. Steinhardt and
N. Turok,  Phys. Rev. {\bf D64}, 123522, (2001); P. J. Steinhardt,
N. Turok,  Phys. Rev. {\bf D65}, 126003, (2002).
\bibitem{Ries98} A.~G. Riess et al., Astron. J. {\bf 116}, 1009, (1998);
 S. Perlmutter et al., Astrophys. J. {\bf 517}, 565, (1999); P. de Bernardis et al.,
 Nature {\bf 404}, 955, (2000); C.~L. Bennett et al., Astrophys. J. Suppl. {\bf 148},
 1 (2003); M. Tegmark et al., astro-ph/0310723.

\bibitem{Rasanen}S. Rasanen,  Nucl. Phys. {\bf B626}, 183, (2002).
\bibitem{maro}A. L. Maroto, Nucl. Phys. {\bf B653}, 109, (2003).
\bibitem{Cand77} P. Candelas and D. Deutsch, Proc. Roy. Soc.
Lond. A {\bf 354}, 79, (1977).
\bibitem{sahrin}R. M. Avagyan, A. A. Saharian, A. H. Yeranyan, Phys. Rev. {\bf D66}, 085023,
 (2002).
 \bibitem{set5}M. R. Setare and R. Mansouri. Class. Quant. Grav. {\bf
 18}, 2659, (2001).
 \bibitem{set6}A. A. Saharian, M. R. Setare, Phys. Lett. {\bf B552},
119, (2003).
\bibitem{Saha03emt}A. A. Saharian, hep-th/0308108.
\end{thebibliography}
\end{document}